\documentclass[seceq]{ptptex}

\def\SU{\mathop{\rm SU}}
\def\U{\mathop{\rm U}}
\def\tr{\mathop{\rm tr}}

\title{
Large $N$ vector quantum mechanics\\
and bubbling supertube solutions%
}

\author{
Yosuke \textsc{Imamura}\thanks{E-mail: \tt imamura@hep-th.phys.s.u-tokyo.ac.jp}%
}

\inst{
Department of Physics, University of Tokyo, Tokyo 113-0033, Japan
}

\abst{
We propose a large $N$ vector quantum mechanics as the theory
describing a D-particle probe in bubbling supertube solutions.
We compute the effective action of this quantum mechanics
and show that
it coincides with the D-particle action in a certain decoupling limit,
up to quadratic order in the velocity.
The angular momentum of the D-particle,
including the contribution of the Poynting vector,
is reproduced as the vacuum expectation value of the $\SU(2)_R$ current.
}

\begin{document}

\maketitle

\section{Introduction}
Recently, many BPS solutions in various supergravity theories
have been constructed
for the purpose of
obtaining new examples of AdS/CFT correspondence.
Lin, Lunin and Maldacena\cite{LLM} constructed a large class of smooth solutions
to type IIB supergravity and M-theory,
which are called bubbling solutions.
Each of them is characterized by a two-dimensional plane
consisting of black and white regions, which is called
a `droplet.'
In the type IIB case, these solutions are dual to a free fermion system,
which describes the dynamics of the BPS sector of N=4 Yang-Mills
theory in ${\bf S}^3$\cite{LLM,Corley:2001zk,toy,Caldarelli:2004ig,buchel,suryanarayana,Caldarelli:2004mz,bbfh,sjt,mandal,mini,tt,qg,mr,Silva:2005fa}.
We can identify the droplet with the phase space structure
of the fermion system.
The M-theory bubbling solutions are constructed in Refs. \citen{LLM} and \citen{fivebrane},
and it is shown that they are related
to BPS sectors of several different gauge theories.
Bubbling solutions in M-theory are also studied in Refs. \citen{bsy,spa,toda}.
Different droplets with the same asymptotic forms give
different classical supergravity solutions with the same boundary conditions.
The existence of many smooth
solutions sharing the same asymptotic behavior
sheds new light on the black hole information problem\cite{mathur}.
Some generalizations of bubbling solutions are given in Refs. \citen{snsn,mm,clp,lv,bonisilva}.

Another major advance in the study of black holes
is represented by the theoretical discovery of black rings\cite{EmpRea}.
They are classical solutions in five-dimensional gravity with horizons
of topology ${\bf S}^2\times{\bf S}^1$.
These solutions are important, because
they constitute counterexamples to the conjecture of black hole uniqueness.
Black rings were soon generalized to supersymmetric ones\cite{EEMR,EEMS2,GauGut}.
More general $1/2$ BPS solutions in ${\cal N}=1$ five-dimensional
supergravity, which include supersymmetric black rings as
special cases, were
subsequently constructed in Refs. \citen{GGHPS} and \citen{BW}.

Based on these results,
a new kind of bubbling solution,
which resolves the singularity of the black ring solutions,
is proposed in Refs. \citen{BW0505} and \citen{Berglund:2005vb}.
In Ref. \citen{BW0505}
the solutions are called ``bubbling supertube solutions''.
Bubbling supertube solutions are the subject of this paper.
Before explaining them,
we need to understand the structure of the solutions
constructed in Refs. \citen{GGHPS} and \citen{BW}.

Let us consider $5$-dimensional ${\cal N}=1$ supergravity
with $n_v$ $\U(1)$ vector multiplets.
The metric of the general $1/2$ BPS solution
has the form
\begin{equation}
ds_5^2=-\frac{1}{Z^2}(dt+k)^2+Zds_4^2,
\end{equation}
where $ds_4^2$ is a four-dimensional hyper K\"ahler base manifold.
When the base manifold is of Gibbons-Hawking (GH) type,
the solution can be explicitly represented by $2n_v+4$
harmonic functions\cite{BKW0504}.
In this paper, we consider only the $n_v=2$ case with the specific
Chern-Simons coefficient
$C_{IJK}=|\epsilon_{IJK}|$.
This theory is obtained by ${\bf T}^6$ compactification and
an appropriate truncation of M-theory.
The GH metric of the base manifold
is determined by one of these harmonic functions,
which is referred to as $V$ in Refs. \citen{BKW0504} and \citen{BW0505}
and in this paper, and is given by
\begin{equation}
ds_4=\frac{1}{V}(d\psi+A)^2+Vd\vec y^2.
\label{GHmetric}
\end{equation}
The differential $A$ is the magnetic dual of the function $V$
satisfying $dA=*dV$,
where $*$ is the Hodge dual
in the flat $3$-dimensional space parameterized by $\vec y=(y_1,y_2,y_3)$.
For concreteness,
we choose the function $V$ for an $n$-center solution as
\begin{equation}
V=V_0+\sum_{i=1}^n\frac{N_i}{4\pi|\vec y-\vec y_i|}.
\label{Vintro}
\end{equation}
When we deal with a metric in the form (\ref{GHmetric}),
we ordinarily assume that all
the coefficients $N_i/4\pi$ and
the constant part $V_0$ are non-negative
in order to guarantee the positive definiteness of the hyper K\"ahler metric
(\ref{GHmetric}).
In Refs. \citen{BW0505} and \citen{Berglund:2005vb},
however, it is shown
that this restriction is in fact
not necessary for the solution in Refs. \citen{GGHPS} and \citen{BW}.
We may choose any harmonic function $V$ of the form (\ref{Vintro})
as long as the coefficients of the poles are appropriately quantized.

Using such generalized GH base spaces,
smooth solutions that resolve
the singularities of supertube solutions are constructed
in Refs. \citen{BW0505} and \citen{Berglund:2005vb}.
In this paper, we call them `bubbling supertube solutions', following Ref. \citen{BW0505}.
In these solutions,
each supertube singularity in the original solutions
is replaced by
a pair of GH centers.
Because of the NUT-charge conservation law,
both positive and negative
NUT charges are needed to construct resolved solutions,
and for this reason, the harmonic function $V$ inevitably takes positive and negative values
depending on the coordinates.
Despite the superficial singularity on the submanifold $V=0$,
actually there are no singularities in
either the metric or the gauge fields
if other harmonic functions are chosen appropriately.
Instead, the submanifold $V=0$ turns out to be ergospheres\cite{Berglund:2005vb}
on which the world line of a stationary point particle is light-like.
(We admit orbifold singularities at centers,
because they are not singular in string theory and
are harmless.)

By treating the coordinate $\psi$ in (\ref{GHmetric}) as the $11$-th coordinate,
these solutions can be regarded as classical solutions in ${\cal N}=2$
four-dimensional supergravity,
which is the ${\bf T}^6$ compactification of type IIA string theory
with a certain truncation.
From this point of view,
BPS particles in uncompactified four-dimensional
spacetime can be regarded as D-branes wrapped on
different holomorphic cycles in the ${\bf T}^6$.
Such four-dimensional solutions has been
constructed by Denef et al.\cite{Denef,DenefGR,DenefB}
independently of the five-dimensional solutions.
The relation between four-dimensional solutions and five-dimensional
solutions is discussed in Ref. \citen{Behrndt:2005he}.

The distinguishing properties of the bubbling supertube solutions
(or corresponding four-dimensional solutions) is that
the particles%
\footnote{Here we use the term ``particles'' in the four-dimensional
sense.
From the viewpoint of five-dimensional spacetime,
it means two different kinds of objects: CG centers and rings.}
in a system interact with one another through a
non-trivial potential and form bound states.
In a static bound state, the positions of particles
are restricted by the so-called ``bubble equation''\cite{BW0505,Berglund:2005vb}.
Because the particles in this system are D-branes
wrapped on internal cycles,
it is naturally conjectured
that the bound states can be studied by computing the potential
energy with boundary states by using the relation
$V\sim\int dt\langle B|e^{-tL_0}|B\rangle$.
This, however, is not the case,
because this computation
takes account of only the linear part of
the gravitational interaction and
gives just a Newtonian potential.
There is a well-known theorem (Ernshaw's theorem)
which states that particles interacting through a
Newtonian potential cannot form static stable bound states.
This implies that the non-linear nature of gravity
plays an important role in the formation of
bound states.

The purpose of this paper is to seek a quantum mechanics that
describes the dynamics of these particles.
Unfortunately, we have not succeeded in constructing such a quantum mechanics for
an entire system of bound particles.
In this paper, we focus on one of the particles in the system
and construct a quantum mechanics describing this particle.
We treat the particle that we chose as a probe and the other
particles as the background.
Furthermore,
we assume that the probe particle carries only D-particle charge
for simplicity.

If a D-particle is placed at a generic point in a bubbling supertube solution,
it will be caused to move by gravitational and RR forces.
There are, however, loci on which D-particle can remain still.
These stability loci are in fact the ergospheres
mentioned above\cite{Berglund:2005vb}.
The existence of such stability loci enables us to consider
the theory on a probe D-particle in the backgrounds.
As mentioned above, the non-linear nature of gravity is essential
for the existence of D-particle stability loci.
From the viewpoint of quantum mechanics,
this implies that the $1$-loop effect, which corresponds to the
string cylinder amplitude
and the exchange of free gravitons, is not sufficient to explain the stability of
the D-particle.
Actually, we show that the two-loop quantum correction plays an important role
in the emergence of stable vacua.

The rest of this paper is organized as follows.

In the next section, we study the action of a D-particle
in bubbling supertube solutions.
We first restrict our attention
to a special class of solutions
in which the function $V$ has only one positive pole and the other seven harmonic
functions are constant.
Although these solutions cannot be regarded as a resolved version of any singular
solution, because of the absence of negative poles in $V$,
they have the distinguishing feature of bubbling supertube solutions that
the function $V$ takes both positive and negative values if the constant part
is negative.

This special solution is actually the
supergravity description of coincident D6-branes in
a constant $B$-field background.
By quantizing open strings,
we obtain supersymmetric $\U(1)$ gauge theory with
$N$ chiral multiplets, where $N$ denotes the number of D6-branes.
When $N$ is large,
the quantum mechanics becomes large $N$ vector quantum mechanics.
In \S\ref{vqm.sec}
we find nice agreement between the D-particle action
and the effective action of this quantum mechanics
in a certain decoupling limit.
We also find that the angular momentum of fluxes induced by the probe
is reproduced as a quantum correction to the $\SU(2)_R$ current.

In \S\ref{gen.sec} we
investigate the generalization to multicenter solutions which include both
positive and negative GH centers.
We propose a quantum mechanics which reproduces
the D-particle action as the effective action.
Finally, we conclude in \S\ref{conc.sec}.

We use the following conventions in this paper.
\begin{itemize}
\item
$2\pi\alpha'^{1/2}=1$ is chosen to be our unit of length.
With this convention, the string tension is $2\pi$, and
the mass and the charge of a D-particle
are $2\pi/g_{\rm str}$ and $2\pi$, respectively.
\item
We normalize gauge fields in such a way that fluxes obtained
by integration over closed surfaces are
quantized as integers.
\item The period of the ${\bf S}^1$ coordinate $\psi$
on the GH base space (\ref{GHmetric}) is $1$.
In this case, NUT charges $N_i$ in (\ref{Vintro}) must be integers.
\end{itemize}

\noindent
{\bf Note Added:}

After submitting this paper to the arXiv,
I was informed of Ref. \citen{Denef:2002ru},
in which a quantum mechanical description for
BPS black hole bound states in ${\cal N}=2$ supergravity
is studied. In particular, the coincidence
of the quantum moduli-space of the quantum
mechanical system and the stability
loci for particles in supergravity classical
solutions is demonstrated there
for sets of charged particles
that are more general than those we consider in this paper.
In this paper, we show that not only the
stability loci but also the D-particle potential
in a specific class of solutions
is reproduced as the effective potential
of the quantum mechanics
in a certain decoupling limit.

\section{D-particle probe in bubbling supertube solutions}\label{sg.sec}
\subsection{$1/2$ BPS solutions}
The most general $1/2$ BPS solutions
in $5$-dimensional ${\cal N}=1$ supergravity
with GH base spaces
are constructed in Ref. \citen{BKW0504}.
They are described with
eight harmonic functions, $V$, $K^I$, $L_I$, $M$,
associated with the NUT charge, three M5 charges, three M2 charges,
and the KK momentum, respectively.
It is convenient for our purposes
to rewrite the solutions
in terms of type IIA string language.
The $11$-dimensional metric $ds_{11}^2$ and the $3$-form potential
$A_3$ are related to the string metric $ds_{10}^2$,
the dilaton $\phi$,
the NS $2$-form $B_2$,
the RR $1$-form $C_1$ and the RR $3$-form $C_3$
as
\begin{equation}
ds_{11}^2=e^{-(2/3)\phi}ds_{10}^2+e^{(4/3)\phi}(d\psi+C_1)^2,\quad
A_3=C_3+B_2\wedge(d\psi+C_1).
\label{general11iia}
\end{equation}
According to these relations, the solutions in Ref. \citen{BKW0504}
can be rewritten as
\begin{eqnarray}
ds_{10}^2&=&-\frac{1}{\sqrt{Q}}(dt+\omega)^2+\sqrt{Q}d\vec y^2
           +\sum_{I=1}^3\frac{\sqrt{Q}}{Z_IV}(dz_{2I-1}^2+dz_{2I}^2),\\
C_1&=&A-\frac{V^2\mu}{Q}(dt+\omega),\\
C_3&=&
\sum_{I=1}^3
\left[\frac{V}{Q}\left(\mu K^I-\frac{Z^3}{Z_I}\right)(dt+\omega)+\xi^I\right]
\wedge dz_{2I-1}\wedge dz_{2I},\\
B_2&=&\sum_{I=1}^3
\left(\frac{K^I}{V}-\frac{\mu}{Z_I}\right)dz_{2I-1}\wedge dz_{2I},
\label{b2harmo}\\
e^\phi&=&\frac{Q^{3/4}}{(ZV)^{3/2}},
\end{eqnarray}
where $z_k$ denotes coordinates in ${\bf T}^6$.
The quantities $Z_I$, $\mu$ and $Q$ are rational functions of the harmonic functions
defined by
\begin{eqnarray}
Z_I&=&L_I+\frac{1}{2}C_{IJK}\frac{K^JK^K}{V},\\
\mu&=&M+\frac{K^IL_I}{2V}+\frac{K^1K^2K^3}{V^2},\\
Q&=&Z^3V-\mu^2V^2,
\end{eqnarray}
and $Z$ is the geometric average $Z=(Z_1Z_2Z_3)^{1/3}$.
The differentials $\omega$ and $\xi^I$ are obtained by solving
certain linear differential
equations\cite{BKW0504,BW0505,Berglund:2005vb}.

For simplicity,
let us first consider solutions in which
the harmonic function $V$ has only one positive pole and
the other harmonic functions are constant.
We study general solutions in \S\ref{gen.sec}.
The constant $M$ must be zero for the solution to be regular.
To fix the other constants, $K^I$ and $L_I$, 
we impose the boundary conditions
\begin{equation}
\lim_{r\rightarrow\infty}
Q=1,\quad
\lim_{r\rightarrow\infty}
VZ_I=g_{\rm str}^{-2/3},
\end{equation}
where $r\equiv |\vec y|$.
These two imply
that the four-dimensional part of the metric
becomes flat Minkowski, $-dt^2+d\vec y^2$,
and $e^\phi$ goes to $g_{\rm str}$ at infinity.
The following choice of the harmonic functions
satisfies these conditions:
\begin{equation}
V=\frac{v_0}{g_{\rm str}}+\frac{N}{4\pi r},
\label{eq15}
\end{equation}
\begin{equation}
K^I=g_{\rm str}^{-1/3}k^I,\quad
L_I=\frac{g_{\rm str}^{1/3}}{v_0}\left(1-\frac{1}{2}C_{IJK}k^Jk^K\right),\quad
M=0,
\label{hamo}
\end{equation}
where $v_0$ and $k^I$ are parameters satisfying
\begin{equation}
v_0^2=1-\frac{1}{4}(k_1+k_2+k_3-k_1k_2k_3)^2.
\label{v0kkk}
\end{equation}
With this choice of the functions, the asymptotic form of the metric is
\begin{equation}
ds_{10}^2(r\rightarrow\infty)=-dt^2+d\vec y^2+g_{\rm str}^{2/3}dz_i^2.
\end{equation}
We assume that the size of the ${\bf T}^6$,
which is determined by the periods of the coordinates $z_k$,
is much larger than the string scale,
in order to make the wrapped D6-branes,
which we treat as a static background, sufficiently heavy.

In fact, the harmonic functions
(\ref{eq15}) and (\ref{hamo})
give the supergravity description of $N$ coincident D6-branes in a constant $B$-field.
The asymptotic value of the $B$-field is
\begin{equation}
B_2(r\rightarrow\infty)=-\sum_{I=1}^3 b_Ie^{2I-1}\wedge e^{2I},
\label{badef}
\end{equation}
where $e^k=g_{\rm str}^{1/3}dz_k$ is the vielbein in the ${\bf T}^6$,
and the three parameters $b_I$ are related to $v_0$ and $k_I$ as
\begin{equation}
b_I=\frac{1}{2v_0}(k_1+k_2+k_3-k_1k_2k_3-2k_I).\label{bbbkkk}
\end{equation}
In order to make the definition of the parameter $b_I$ unambiguous,
we have to specify the gauge choice for $B_2$.
One way to do this is to specify the gauge field $F_2$ on the D6-branes.
In the five-dimensional solutions
we can define the gauge field on the D6-branes by the coupling
with an M2-brane wrapped on a non-compact $2$-cycle,
and we can show that $F_2=0$ for the classical solution
given by (\ref{eq15}) and (\ref{hamo}).

The relations (\ref{v0kkk}) and (\ref{bbbkkk}) are solved
with respect to $k_I$ and $v_0$ to give the solution
\begin{equation}
v_0=\sin\xi,\quad
k_I=\frac{\cos(\beta_I+\xi)}{\cos\beta_I},
\label{v0kisol}
\end{equation}
where the angles $\beta_I$ and $\xi$ are defined by
\begin{equation}
\beta_I=\tan^{-1}b_I,\quad
\xi=\frac{\pi}{2}-\beta_1-\beta_2-\beta_3.
\end{equation}
For definiteness, we assume $0\leq\beta_I\leq\pi/2$.

\subsection{D-particle in the background}
Let us expand the D-particle effective action
in bubbling supertube solutions as
\begin{equation}
L_{\rm D0}=L_0+L_1+L_2+\cdots,\label{ld0}
\end{equation}
where the subscripts represent the power of the velocity $\dot y^i$ in each term.
By substituting this solution
into the DBI and CS actions,
we obtain
\begin{eqnarray}
-V_{\rm D0}=L_0&=&-2\pi\left(\frac{\sqrt{-g_{tt}}}{e^\phi}+C_t\right)
=\frac{2\pi V^2}{\sqrt{V^3Z^3}+V^2\mu}
,\label{l0}\\
L_1&=&-2\pi C_i \dot y^i=-2\pi A_i\dot y^i,\label{lf}\\
L_2&=&2\pi\frac{g_{ij}}{2e^\phi\sqrt{-g_{tt}}}\dot y^i\dot y^j
=\pi(VZ)^{3/2}|\dot y^m|^2.\label{l2}
\end{eqnarray}

Let us consider the potential term
$V_{\rm D0}\equiv -L_0$ first.
Because $VZ$ and $V^2\mu$ are positive and regular for regular solutions,
the minima of this potential are given by $V=0$.
From the viewpoint of M-theory,
the condition $V=0$ gives ergospheres,
which are defined as submanifolds on which
the world-line of a stationary point particle is light-like.
This can be easily confirmed by considering
the eleven-dimensional line element for a stationary particle,
\begin{equation}
ds_{11}^2
=-e^{-(2/3)\phi}\frac{dt^2}{\sqrt{Q}}+e^{(4/3)\phi}(C_t dt)^2
=-\frac{1}{Z^2}dt^2
=-\frac{V^2}{(ZV)^2}dt^2.
\end{equation}
Because $(VZ)^2$ is positive definite and finite,
ergospheres are given by $V=0$.

For simplicity,
let us restrict our attention to single-center solutions
described by the harmonic functions (\ref{eq15}) and (\ref{hamo}).
The potential $V_{\rm D0}$ for these single-center solutions depends
on the parameters $g_{\rm str}$, $N$ and $\beta_I$, and the radial coordinate $r$.
It is invariant under
the replacement $(r,N)\rightarrow(\alpha r,\alpha N)$.
Under another replacement $(g_{\rm str},N)\rightarrow(\alpha g_{\rm str},\alpha^{-1}N)$,
the potential is rescaled as $V_{\rm D0}\rightarrow\alpha^{-1}V_{\rm D0}$.
These two facts partially determine the functional form of the
potential as
\begin{equation}
V_{\rm D0}=\frac{1}{g_{\rm str}}f\left(\frac{Ng_{\rm str}}{r},\beta_I\right).
\end{equation}
The potential $V_{\rm D0}$ depends on the three angles $\beta_I$ through the four parameters $v_0$ and $k_I$,
which are related to each other as in (\ref{v0kkk}).
If this constraint were absent and $v_0$ and $k_I$ were four independent parameters,
the potential would be rescaled as $V_{\rm D0}\rightarrow\alpha V_{\rm D0}$
through the replacement $(v_0,g_{\rm str})\rightarrow(\alpha v_0,\alpha g_{\rm str})$.
This implies that the potential can be written in the form
\begin{equation}
V_{\rm D0}=\frac{v_0^2}{g_{\rm str}}g(\rho,k_I(v_0,\beta_{I'})),\quad
\rho=\frac{v_0 r}{g_{\rm str}N}.
\label{frhok}
\end{equation}
Instead of the three angles $\beta_I$ ($I=1,2,3$),
we choose $v_0=\sin\xi$ and $\beta_{I'}$ ($I'=1,2$) as the three independent variables.
In the next section,
we compare this potential with the effective potential
of a certain supersymmetric quantum mechanics.
We determine the relation between the parameters $g_{\rm str}$, $N$ and $v_0$
and a coupling constant, the number of chiral multiplets,
and FI parameter, respectively.
However, there are no quantities that correspond to the $\beta_{I'}$.
We decouple these unwanted parameters by taking the small $\xi$ limit
as follows.
(It may be possible to introduce extra parameters
corresponding to $\beta_{I'}$
in our quantum mechanics.
However, we do not discuss this possibility here.)
Let us expand the function $g$
in (\ref{frhok})
with respect to $v_0$ as
\begin{equation}
g(\rho,k_I(v_0,\beta_{I'}))
=\sum_{n=0}^\infty g_n(\rho,\beta_{I'})v_0^n.\label{gexp}
\end{equation}
Because
the derivative of $k_I$  with respect to $\beta_I$ produces an extra factor of $v_0$,
\begin{equation}
\frac{\partial k_I}{\partial\beta_I}=-\frac{v_0}{\cos^2\beta_I},
\end{equation}
the $\beta_{I'}$ dependent terms in $g$ have at least one factor of $v_0$,
and the leading term, $g_0$, on the right-hand side
of (\ref{gexp}) is independent of $\beta_{I'}$.
In this paper,
we focus only on the leading term, $g_0$, in the $v_0$ expansion.
In other words, we take the following small $\xi$ limit
in order to decouple the unwanted parameters $\beta_{I'}$:
\begin{equation}
v_0=\sin\xi\rightarrow 0,\quad\mbox{with}\quad
\rho=\frac{v_0r}{g_{\rm str}N}\quad\mbox{fixed}.
\label{smallxi}
\end{equation}
In this limit,
both the terms $\sqrt{V^3Z^3}$ and $V^2\mu$ in the denominator of the
right-hand side of (\ref{l0}) approach $1/g_{\rm str}$, and the leading term of the D-particle potential in the $\xi$ expansion is
\begin{equation}
V_{\rm D0}(r)
=\frac{2\pi g_{\rm str}}{2}V^2
=\frac{2\pi}{2g_{\rm str}}\left(\xi+\frac{Ng_{\rm str}}{4\pi r}\right)^2.
\label{d0potentialc2}
\end{equation}

Let us take the same limit for the velocity dependent terms $L_1$ and $L_2$.
The term $L_1$ is linear in the velocity, and it represents
the Lorentz force due to the background RR $1$-form potential.
$A_i$ in (\ref{lf}) is the vector potential for a monopole in the $\vec y$ space.
The monopole is located at $\vec y=0$, and its charge is $N$.
This term is, in a sense, topological, and
its form does not change in the
limit (\ref{smallxi}).
The coefficient $(VZ)^{3/2}$ of the kinetic term $L_2$ in (\ref{l2}) becomes
the constant $1/g_{\rm str}$ in the small $\xi$ limit.

Summing $L_0$, $L_1$ and $L_2$,
we obtain the D-particle effective Lagrangian in the small $\xi$ limit,
\begin{equation}
L_{\rm D0}=\frac{2\pi}{2g_{\rm str}}|\dot{\vec y}|^2-2\pi A_i\dot y^i-\frac{2\pi g_{\rm str}}{2}V^2,
\label{multieffdac}
\end{equation}
up to ${\cal O}(\dot y^3)$.

\section{Large $N$ vector quantum mechanics}\label{vqm.sec}
\subsection{Lagrangian}
To obtain the theory on a D-particle probe,
we treat a classical solution as a D-brane system consisting of
background D6-branes and a probe D-particle in a constant $B$-field,
and quantize open strings in it.
This D0-D6 system becomes BPS
when $\xi=0$\cite{Ohta:1997fr,CIMM,Mihailescu:2000dn,witten,Fujii:2001wp}.

Let us assume that $N$ D6-branes are located at the center, $\vec y=0$ in the
transverse space.
One $\U(1)$ vector multiplet
and three neutral chiral multiplets arise in D0-D0 string modes.
The vector multiplet ${\cal V}$ consists of
a (non-dynamical) gauge field, $A_t$, three scalar fields, $\vec a=(a_1,a_2,a_3)$,
and a two-component fermion, $\chi$.
We ignore the neutral chiral multiplets, because they decouple in the $\U(1)$ case.
As the lowest modes in D0-D6 strings,
$N$ charged chiral multiplets $\Phi_\alpha$ ($\alpha=1,\ldots,N$) arise.
Let $\phi_\alpha$ and $\psi_\alpha$
denote the complex bosons and two-component fermions in $\Phi_\alpha$.
These fields carry the same $\U(1)$ charges, $+1$.
Their masses, obtained through the quantization of open strings, are\cite{CIMM,witten,Fujii:2001wp}
\begin{equation}
m_f^2=(2\pi r)^2,\quad
m_b^2=2\pi\xi+(2\pi r)^2,
\label{mass06}
\end{equation}
where $r$ is the distance between the D-particle and the D6-branes.
These masses become equal on the supersymmetric locus $\xi=0$
in the parameter space.

The tree-level Lagrangian for these multiplets, which is obtained through
the dimensional reduction
of the four-dimensional ${\cal N}=1$ Lagrangian, is
\begin{eqnarray}
L_{\rm tree}
&=&\frac{1}{g_{\rm qm}^2}
\left[\left(\int d^2\theta W^2+\mbox{c.c.}\right)+\int d^4\theta \zeta{\cal V}
+\sum_{\alpha=1}^N\int d^4\theta \Phi_\alpha^\ast e^{\cal V}\Phi_\alpha\right]
\nonumber\\
&=&\frac{1}{g_{\rm qm}^2}
\Big[\frac{1}{2}(\partial_t\vec a)^2
    +\chi^\dagger\partial_t\chi
    -\frac{1}{2}D^2
\nonumber\\&&\quad
    +\sum_{\alpha=1}^N\Big(
     |D_t\phi_\alpha|^2
    -\vec a^2|\phi_\alpha|^2
    +\psi_\alpha^\dagger D_t\psi_\alpha
\nonumber\\&&\quad\quad\quad
    +\psi_\alpha^\dagger\vec\sigma\cdot\vec a\psi_\alpha
    +\phi_\alpha^\dagger\chi\psi_\alpha
             +\phi_\alpha\chi^\dagger\psi_\alpha^\dagger\Big)\Big],
\label{treepot}
\end{eqnarray}
where $\vec\sigma=(\sigma_x,\sigma_y,\sigma_z)$ are the Pauli matrices,
and $D$ is the auxiliary field in the vector multiplet.
The equation of motion for the auxiliary field has already been solved
in the right-most expression in (\ref{treepot}),
and $D$ in (\ref{treepot}) is given by
\begin{equation}
D=\zeta+\sum_{\alpha=1}^{N} |\phi_\alpha|^2.
\label{auxD}
\end{equation}
The Lagrangian (\ref{treepot}) is invariant with respect to an $\SU(2)_R$
transforming
$\vec a$, $\chi$, and $\psi_\alpha$
as $\bf 3$, $\bf 2$, and $\bf 2$, respectively.
There is no superpotential because all the chiral multiplets carry
the same charge.
The classical masses of the bosons $\phi_\alpha$ and fermions $\psi_\alpha$
read off from the Lagrangian are
\begin{equation}
m_\psi^2=|\vec a|^2,\quad
m_\phi^2=|\vec a|^2+\zeta.
\label{classicalmass}
\end{equation}
The relation between the coupling constants $g_{\rm qm}$ and $g_{\rm str}$
is determined by comparing the DBI action of the D-particle and the kinetic term
in (\ref{treepot}).
We also obtain other relations from comparison
of (\ref{mass06}) and (\ref{classicalmass}).
Specifically, we have
\begin{equation}
g_{\rm qm}^2=2\pi g_{\rm str},\quad
\vec a=2\pi\vec y,\quad
\zeta=2\pi\xi,
\label{paramcorr}
\end{equation}
where $\vec y$ is the position of the probe D-particle.

The classical supersymmetric vacuum conditions are
\begin{equation}
\vec a|\phi_\alpha|=0,\quad
D=0.
\label{classicalvacua}
\end{equation}
When $\zeta<0$, the chiral multiplets acquire the non-vanishing
vacuum expectation value $|\phi_\alpha|=\sqrt{|\zeta|}$,
and we have $\vec a=0$.
Because the overall phase of $\phi_\alpha$ is a gauge degree of freedom,
the moduli space is ${\bf CP}^{N-1}$.
When $\zeta=0$, all the $\phi_\alpha$ must be zero,
and the moduli space is ${\bf R}^3$ parameterized by $\vec a$.
When $\zeta>0$, there is no supersymmetric vacuum.

The quantum mechanics represented by the Lagrangian in (\ref{treepot}) was first suggested in Ref. \citen{witten}
in connection to the D0-D6 system,
and the relation between its classical vacuum structure
and the behavior of D0-D6 system was also discussed in Ref. \citen{witten}.
In what follows, we see that quantum corrections in our quantum mechanics
reproduce the action of a D-particle in the supergravity solutions.

\subsection{One-loop and two-loop corrections}
In order to demonstrate the validity of the treatment of a D-particle as a probe,
we assume that the number $N$ of D6-branes is much larger than
$1$, the number of the probe D-particle,
and study only the leading term
of the $1/N$ expansion.
More precisely, we take the following large $N$ limit:
\begin{equation}
N\rightarrow\infty,\quad
\mbox{with}\quad
\lambda\equiv Ng_{\rm qm}^2\quad\mbox{fixed}.
\label{largeN}
\end{equation}
As is easily checked with the Feynman rules obtained from the Lagrangian
(\ref{treepot}), no diagrams containing $\vec a$ or $\chi$ as internal lines
appear in the leading correction of the $1/N$ expansion.
This implies that
these fields behave like classical external fields
in the large $N$ limit.
We define the effective action
as a functional of the classical external fields $\vec a$ and $\chi$.
The fermion $\chi$ is set to zero in our analysis.

Corresponding to (\ref{smallxi}) on the supergravity side,
we take the small $\zeta$ limit
\begin{equation}
\frac{\zeta}{|\vec a|^2}\rightarrow 0,\quad\mbox{with}\quad
\frac{\lambda}{\zeta|\vec a|}\quad\mbox{fixed},
\label{smzeta}
\end{equation}
in addition to
the large $N$ limit given in (\ref{largeN}).
As we show below,
the effective potential is two-loop exact
in this small $\zeta$ limit.
It is known that
in the context of the M(atrix) theory,
the leading and sub-leading potential
of the D0-D6 system
can be reproduced as $1$-loop\cite{pierre,BISY,Lif,KVK,DM} and two-loop\cite{branco,dhar} corrections,
respectively.
In this section, we confirm that the quantum mechanics proposed above
also reproduces
the D-particle effective action given in \S\ref{sg.sec}.

Let us decompose the effective potential $V_{\rm eff}(\vec a)$ into three parts
$V_{\rm tree}$, $V_f$ and $V_b$.
$V_{\rm tree}$ is the tree-level potential:
\begin{equation}
V_{\rm tree}(\vec a)=\frac{1}{2g_{\rm qm}^2}\zeta^2.
\label{Vtree}
\end{equation}
We set $\phi_\alpha=0$, because
$\phi_\alpha$ cannot acquire a non-vanishing vacuum expectation value,
guaranteed by Coleman's theorem.
$V_f$ represents quantum correction, including fermions, $\psi_\alpha$.
At leading order in the $1/N$ expansion
of the effective action, the only diagram, including
fermions $\psi_\alpha$ is the $1$-loop diagram which gives the
contribution
\begin{equation}
V_f(\vec a)
=-N\int\frac{dk}{2\pi}\log(k^2+m_\psi^2)
   =-N\left(m_\psi+\frac{2}{\pi}\Lambda(\log\Lambda-1)\right),
\end{equation}
where $\Lambda$ is a momentum cut off.
The rest of the quantum corrections, which are collectively denoted by $V_b$,
consist of the contribution of the scalar fields $\phi_\alpha$.
Although there are an infinite number of loop diagram
of scalar fields $\phi_\alpha$,
the large $N$ assumption makes it quite easy to compute them.
The loop momentum integrals in any multi-loop diagrams are
factorized and are easily carried out.
For example, the only two-loop diagram is the ``8''-shaped diagram,
which is essentially the square of a one-loop diagram.
The one-loop and two-loop contributions to $V_b$ are given by
\begin{eqnarray}
V_{b(1-loop)}(\vec a)
&=&N\int\frac{dk}{2\pi}\log(k^2+m_\phi^2)=N\left(m_\phi+\frac{2}{\pi}\Lambda(\log\Lambda-1)\right),\\
V_{b(2-loop)}(\vec a)
&=&\frac{g_{\rm qm}^2}{2}
    \left(N\int\frac{dk}{2\pi}\frac{1}{k^2+m_\phi^2}\right)^2
    =\frac{g_{\rm qm}^2N^2}{8m_\phi^2}.
\end{eqnarray}
Because of the supersymmetry,
the divergent terms in $V_{b(1-loop)}$ and $V_f$ cancel.
Summing the contributions up to the two-loop order,
we obtain
\begin{equation}
V_{\rm eff}(\vec a)=V_{\rm tree}(\vec a)+V_f(\vec a)+V_b(\vec a)
=\frac{N}{2\lambda}\left(\zeta
+\frac{\lambda}{2|\vec a|}\right)^2.
\label{vlarger}
\end{equation}

In the small $\zeta$ limit,
the two-loop effective potential obtained above is
exact,
and there is no higher-loop contribution to $V_{\rm eff}$.
This is shown as follows.
Because $V_f$ is one-loop exact in the large $N$ limit,
the potential higher-loop contributions are
multi-loop diagrams of the scalar fields $\phi_\alpha$.
They depend on $\zeta$ and $|\vec a|$ only through the bare scalar mass $m_\phi=\sqrt{|\vec a|^2+\zeta}$.
From dimensional analysis, it is found that the $L$-loop contribution is $\lambda^L/m_\phi^{3L}$, up to a numerical constant.
This is expanded with respect to $\zeta$ as
\begin{equation}
V_{b(L-loop)}
=\sum_{P=0}^\infty\frac{c_{L,P}}{g_{\rm qm}^2}
\left(\frac{\lambda}{\zeta|\vec a|}\right)^L\left(\frac{\zeta}{|\vec a|^2}\right)^{P+L},
\end{equation}
where $c_{L,P}$ in the equation represents numerical
coefficients obtained in the loop calculation.
From this expression, it is apparent that any diagrams
with more than $3$ loops give terms of higher order in $\zeta/|\vec a|^2$,
which should be ignored in the small $\zeta$ limit.
Thus, the effective potential (\ref{vlarger}) is exact in the large $N$, small $\zeta$ limit.

If we rewrite the effective potential (\ref{vlarger})
in terms of variables on the supergravity side
according to the parameter correspondence
(\ref{paramcorr}),
we find it
coincides with the D-particle potential (\ref{d0potentialc2}).

It is useful to note that in the large $N$, small $\zeta$ limit,
the relation
\begin{equation}
V_{\rm eff}=\frac{1}{2g_{\rm qm}^2}\langle D\rangle^2
\label{DV2}
\end{equation}
holds,
where $\langle D\rangle$ is the vacuum expectation value of the auxiliary field
(\ref{auxD}),
which is one-loop exact
in the large $N$, small $\zeta$ limit.
The following relation, somewhat simpler than (\ref{DV2}), also holds:
\begin{equation}
V=\frac{\langle D\rangle}{g_{\rm qm}^2}.
\label{VandD}
\end{equation}

To this point, we have focused on the effective potential.
To complete the comparison between the D-particle action (\ref{ld0})
and the effective action of our quantum mechanics,
let us check the coincidence of the velocity dependent terms $L_2$ and $L_1$.

The fermion one-loop correction
to the two-point function
takes the form $\Gamma^{(2)}=\delta a_i(p)M_{ij}\delta a_j(-p)$
with
\begin{eqnarray}
M_{ij}&=&-N\int\frac{dk}{2\pi}
\tr\left(
\sigma_i\frac{i}{k+p/2+i\vec a\cdot\vec\sigma}
\sigma_j\frac{i}{k-p/2+i\vec a\cdot\vec\sigma}
\right)
\nonumber\\
&=&-\frac{\partial^2 V_f}{\partial a_i\partial a_j}
+N\frac{\epsilon_{ijk}a_k}{2m_f^3}p
-N\frac{a^2\delta_{ij}-a_ia_j}{4m_f^5}p^2
+{\cal O}(p^3).
\label{mij}
\end{eqnarray}
The first term here is the second derivative of $-V_f(\vec a)$
and is independent of $p$.
The second term in (\ref{mij}) is linear in $p$.
This term implies the existence of the
parity-violating term in the effective action given by
\begin{equation}
\Gamma
=\int\frac{N}{2m_f^3}\epsilon_{ijk}a_i\delta a_j\partial_t\delta a_k dt
=2\pi N\int d\vec a\cdot\vec A,
\end{equation}
where $\vec A$ is the monopole potential in the $\vec a$ space,
which is given by
\begin{equation}
\vec A\sim \frac{1}{4\pi |\vec a|^3}\vec a\times\delta \vec a,
\end{equation}
in the vicinity of $\vec a$.
This reproduces the Lorentz force term $L_1$.
The appearance of this term in the one-loop correction
is also found in Ref. \citen{lorentz} in the context of
M(atrix) theory.

The third term in (\ref{mij}) yields a wave function renormalization of order $\lambda/|\vec a|^3$.
The loop diagrams of the scalar fields $\phi_\alpha$ also yield
wave function renormalization
of the same order of magnitude.
These corrections vanish
in the small $\zeta$ limit (\ref{smzeta}),
and the kinetic term of $\vec a$ is not corrected in this limit.
This is consistent with the D-particle kinetic term
with a constant coefficient appearing in (\ref{multieffdac}).

We have thus confirmed that the quantum mechanics represented by (\ref{treepot})
reproduces the D-particle action (\ref{multieffdac})
as the effective action.
Before ending this section, we give one more example of the correspondence
between a classical quantity in the supergravity
and a loop correction in our quantum mechanics.
Let us consider the $\SU(2)_R$ symmetry,
which rotates the $\vec a$ space.
It transforms
the fermions $\chi$ and $\psi_\alpha$ as doublets,
and the current is
\begin{equation}
\vec j=\frac{1}{g_{\rm qm}^2}(
\vec a\times\partial_t\vec a
+\chi^\dagger\vec\sigma\chi
+\psi_\alpha^\dagger\vec\sigma\psi_\alpha
).
\label{jvec}
\end{equation}
Because we treat $\vec a$ and $\chi$ as background classical fields
describing the classical motion of the D-particle,
the first two terms just give the orbital angular momentum of the D-particle.
In addition, the third term has
a non-vanishing vacuum expectation value,
due to the fermion one-loop correction:
\begin{equation}
\langle\vec j\rangle
=N\int\frac{dk}{2\pi}\tr\left(\sigma_z\frac{i}{k+i\vec a\cdot\vec\sigma}\right)
=N\frac{\vec a}{|\vec a|}.
\label{su2vev}
\end{equation}
This should also be identified with the angular momentum of the probe D-particle.
It is well known that systems consisting of mutually non-local charges
can have non-vanishing angular momenta,
due to the non-vanishing Poynting vector.
In our case, it
can be evaluated, for example,
by considering the asymptotic behavior of the differential $\omega$
perturbed by the probe D-particle.
We obtain
\begin{equation}
\vec J=N\frac{\vec y}{r},
\end{equation}
and this coincides with (\ref{su2vev}).

\section{Generalization to multicenter solutions}\label{gen.sec}
To this point, we have only treated single center solutions
and the quantum mechanics with $N$ identical chiral multiplets.
As demonstrated below, however,
it is possible to generalize this quantum mechanics so that
its effective action
reproduces the D-particle effective action in
arbitrary bubbling solutions.

In the previous section, we introduced the parameters $\beta_I$ as those determining the background $B$-field.
There, the lower-dimensional brane charge dissolved in D6-branes is induced by the Chern-Simons term in the
D6-brane action.
This, however, cannot be done in general cases, in which both D6-branes and anti-D6-branes
with different lower dimensional brane charges exist.
In such cases, it is more convenient to regard the parameters $\beta_I$
as that determining the gauge fields on D6-branes, not the background.
For this reason, we perform a $B$-field gauge transformation
so that the asymptotic value of the $B$-field vanishes.
For the bubbling solutions,
$B$-field transformations amount to the following transformations
of the harmonic functions\cite{BKW0504,BW0505}:
\begin{eqnarray}
K^I&\rightarrow& K^I+c^IV,\nonumber\\
L_I&\rightarrow& L_I-C_{IJK}c^JK^K-\frac{1}{2}C_{IJK}c^Jc^KV,\nonumber\\
M&\rightarrow& M-\frac{1}{2}c^IL_I+\frac{1}{12}C_{IJK}(Vc^Ic^Jc^K+3c^Ic^JK^K).
\end{eqnarray}
To make the asymptotic value of $B_2$ vanish, we should
carry out this transformation with parameters $c^I=g_{\rm str}^{2/3}b_I$.
Doing so, we obtain
\begin{eqnarray}
V&=&\frac{1}{g_{\rm str}}\left(\sin\xi+\frac{Ng_{\rm str}}{4\pi r}\right),\nonumber\\
K^I&=&\frac{1}{g_{\rm str}^{1/3}}\left(\cos\xi+b_I\frac{Ng_{\rm str}}{4\pi r}\right),\nonumber\\
L_I&=&g^{1/3}_{\rm str}\left(\sin\xi-\frac{1}{2}C_{IJK}b_Jb_K\frac{Ng_{\rm str}}{4\pi r}\right),\nonumber\\
M&=&-\frac{g_{\rm str}}{2}\left(\cos\xi
    -b_1b_2b_3\frac{Ng_{\rm str}}{4\pi r}\right).
\end{eqnarray}
In this gauge, the constant parts of the harmonic functions depend only on $\xi$,
while the terms proportional to $1/r$ contain $b_I$ separately.
Here, let us treat the parameters $\xi$ and $\beta_I$ as independent.
Then we regard $\beta_I$ as parameters for the D6-branes and $\xi$ as that for the background.
The relation (\ref{v0kisol}) among $\xi$ and $\beta_I$ is obtained
as the bubble equation\cite{BW0505,Berglund:2005vb} for single center solutions.

Because the parameters $\beta_I$ are contained only in the pole terms,
the solution can easily be generalized to $n$-center solutions
by superposing single-center solutions as follows:
\begin{eqnarray}
V&=&\frac{1}{g_{\rm str}}
\left(\sin\xi+\sum_{i=1}^n\frac{N_ig_{\rm str}}{4\pi|\vec y-\vec y_i|}\right),\nonumber\\
K^I&=&\frac{1}{g_{\rm str}^{1/3}}
\left(\cos\xi+\sum_{i=1}^n b^i_I\frac{N_ig_{\rm str}}{4\pi|\vec y-\vec y_i|}\right),\nonumber\\
L_I&=&g^{1/3}_{\rm str}
\left(\sin\xi-\frac{1}{2}\sum_{i=1}^n C_{IJK}b_J^ib_K^i\frac{N_ig_{\rm str}}{4\pi|\vec y-\vec y_i|}\right),\nonumber\\
M&=&-\frac{g_{\rm str}}{2}\left(\cos\xi
    -\sum_{i=1}^n b_1^ib_2^ib_3^i\frac{N_ig_{\rm str}}{4\pi|\vec y-\vec y_i|}\right).
\label{vklmmulti}
\end{eqnarray}
We assign different $\beta_I^i$ to each pole labeled by $i=1,\ldots,n$.
Substituting these harmonic functions into the regularity condition
$\mu(\vec y_i)=0$, we obtain the bubble equation
\begin{equation}
\frac{\sin(\xi_i-\xi)}{\cos\beta^i_1\cos\beta^i_2\cos\beta^i_3}
+\sum_{j\neq i}(b^i_1-b^j_1)(b^i_2-b^j_2)(b^i_3-b^j_3)\frac{N_jg_{\rm str}}{4\pi|\vec y_i-\vec y_j|}=0,
\label{bubbleeq}
\end{equation}
where we define $\xi_i$ for each GH center by
\begin{equation}
\xi_i=\frac{\pi}{2}-\beta_1^i+\beta_2^i+\beta_3^i.
\end{equation}
It is worth noting that the vacuum condition $V=0$ is in fact
a special case of the bubble equation.
Indeed, if we use the index $i=0$ for the probe D-particle,
we obtain $V=0$ as the bubble equation in the limit $\beta_I^0\rightarrow\pi/2$.

We assume that all $\xi_i$ are of the same order as $\xi$,
and
take the following small $\xi$ limit:
\begin{equation}
\xi\rightarrow 0,\quad\mbox{with}\quad
\frac{\xi_i}{\xi},\quad
\frac{\xi\vec y_i}{N_ig_{\rm str}}\quad\mbox{fixed}.
\label{smallxis}
\end{equation}
In this small $\xi$ limit,
the D-particle effective action is given by (\ref{multieffdac}),
with $V$ and $A$ replaced by the harmonic function in (\ref{vklmmulti})
and its magnetic dual, respectively.
This effective Lagrangian does not depend on $\xi^i$.
In the small $\xi$ limit (\ref{smallxis}),
we can always tune these irrelevant parameters $\xi^i$
so that the bubble equation (\ref{bubbleeq}) holds.

Each GH 
center labeled by the index $i$
in (\ref{vklmmulti})
carries a NUT charge $N_i$.
This charge represents the number of corresponding
(anti-)D6-branes.
Because each elementary (anti-)D6-brane gives one chiral multiplet
with charge $+1$ ($-1$),
it is more convenient to label them one by one
when we investigate dual quantum mechanics.
We use the index $\alpha$ for this labeling
and write the harmonic function $V$ as
\begin{equation}
V=\frac{\xi}{g_{\rm str}}+\sum_{\alpha=1}^N\frac{e_\alpha}{4\pi|\vec y-\vec y_\alpha|},
\quad
N=\sum_{i=1}^n|N_i|,
\label{posnegv}
\end{equation}
where $\vec y_\alpha$ and $e_\alpha=\pm1$ are the position
and charge of each (anti-)D6-brane.
We propose the quantum mechanics
consisting of a $\U(1)$ vector multiplet ${\cal V}=(A_t,\vec a,\chi)$ and
$N$ charged chiral multiplets $\Phi_\alpha=(\phi_\alpha,\psi_\alpha)$ with
masses
\begin{equation}
\vec m_\alpha=2\pi \vec y_\alpha
\label{masspole}
\end{equation}
and charges $e_\alpha$
as the theory probing the bubbling solution.
The Lagrangian
for this quantum mechanics is
\begin{eqnarray}
L_{\rm tree}
&=&\frac{1}{g_{\rm qm}^2}
\left[\left(\int d^2\theta W^2+\mbox{c.c.}\right)+\int d^4\theta \zeta{\cal V}
+\sum_{\alpha=1}^N\int d^4\theta \Phi_\alpha^\ast e^{e_\alpha({\cal V}-\theta^\dagger\vec a\cdot\vec\sigma\theta)}\Phi_\alpha\right]
\nonumber\\
&=&\frac{1}{g_{\rm qm}^2}\Big[
     \frac{1}{2}(\partial_t\vec a)^2
    +\chi^\dagger\partial_t\chi
    -\frac{1}{2}D^2
\nonumber\\&&
     +\sum_{\alpha=1}^N\Big(
     |D_t\phi_\alpha|^2
    -|\vec a-\vec m_\alpha|^2|\phi_\alpha|^2
    +\psi_\alpha^\dagger D_t\psi_\alpha
\nonumber\\&&
    +e_\alpha\psi_\alpha^\dagger\vec\sigma\cdot|\vec a-\vec m_\alpha|\psi_\alpha
    +e_\alpha(\phi_\alpha^\dagger\chi\psi_\alpha
    +\phi_\alpha\chi^\dagger\psi_\alpha^\dagger)
    \Big)\Big],
\label{treepot2}
\end{eqnarray}
where the auxiliary field $D$ is given by
\begin{equation}
D=\zeta+\sum_{\alpha=1}^N e_\alpha|\phi_\alpha|^2.
\end{equation}
This Lagrangian is obtained through dimensional reduction from a four-dimensional
${\cal N}=1$ gauge theory.
Although a non-vanishing superpotential is not forbidden by the gauge invariance,
we simply set $W=0$ here.
It may be interesting to seek that wich on the supergravity side
corresponds to turning on a superpotential
in our quantum mechanics.

The computation of the effective action proceed in parallel to that
we did in the previous section.
In the small $\zeta$ limit (\ref{smzeta}), the wave function renormalization
vanishes.
The Lorentz force term in 
(\ref{multieffdac}) is reproduced by the fermion $1$-loop diagram.
In this case, each fermion with mass $\vec m_\alpha$ gives
a monopole at $\vec a=\vec m_\alpha$ in the $\vec a$-space.
Summing up the contributions from all the fermions,
the Lorentz force term is reproduced.

In order to show that the potential term in (\ref{multieffdac}) is
correctly reproduced,
we can use the
relation (\ref{VandD})
instead of computing the effective potential itself.
The expectation value $\langle D\rangle$, which is one-loop exact in the
small $\zeta$ limit, is given by
\begin{equation}
\langle D\rangle
=\zeta+\frac{g_{\rm qm}^2}{2}\sum_{\alpha=1}^{N}
\frac{e_\alpha}{|\vec a-\vec m_\alpha|}.
\label{multiconsis}
\end{equation}
This is identical to the harmonic function $V$ through
the parameter correspondence (\ref{VandD}),
and thus the D-particle potential is correctly reproduced.

The angular momentum of a D-particle in a general bubbling solution
is obtained as the vacuum expectation value of the $\SU(2)_R$ current.
In general,
the background geometry itself may have non-vanishing
angular momentum.
It should be noted that
the $\SU(2)_R$ current gives
only the contribution of the probe, including both 
angular momentum due to the classical motion of the probe
and that due to the Poynting vector induced by the
charge of the probe.

\section{Conclusions}\label{conc.sec}
We proposed a supersymmetric
large $N$ vector quantum mechanics
as the theory describing a probe D-particle in bubbling supertube solutions.

A bubbling supertube solution
can be regarded as a system of D6 and anti-D6 branes
carrying lower-dimensional brane charges.
This solution is parameterized by two parameters, $g_{\rm str}$
and $\xi$, for the asymptotic behavior of fields
and six parameters, $\beta_I^i$ and $\vec y_i$, for every (anti-)D6-brane.
We computed the D-particle effective action in the background and
showed that in the small $\xi$ limit (\ref{smallxis}),
the action does not depend on $\beta_I^i$.
We can always tune these irrelevant parameters
so that the bubble equation holds
for any given positions $\vec y_i$ of the branes.

The quantum mechanics we propose
consists of one $\U(1)$ vector multiplet and $N$ charged chiral multiplets.
Each chiral multiplet corresponds to one (anti-)D6-brane,
and its $\U(1)$ charge is $+1$ for a D6-brane and $-1$ for an anti-D6-brane.
The parameters $g_{\rm str}$, $\xi$, and $\vec y_\alpha$ of the bubbling supertube solution
are mapped to the coupling constant, the FI-parameter, and the bare masses
of the chiral multiplets.
We showed that the D-particle effective action up to quadratic order
in the velocity
is correctly reproduced as the effective action of this quantum mechanics.
Stability loci for the probe D-particle,
which are equivalent to ergospheres in the five-dimensional solution,
correspond to the quantum moduli space of this quantum mechanics.

We also showed that the angular momentum of the probe D-particle
is correctly reproduced in out quantum mechanics as
the one-loop expectation value of the $\SU(2)_R$ charge.

\section*{Acknowledgements}
I would like to thank Y.~Tachikawa for valuable discussions.
This work is supported in part by
a Grant-in-Aid for the Encouragement of Young Scientists
(\#15740140) from the Japan Ministry of Education, Culture, Sports,
Science and Technology.


%

\end{document}